\documentclass[aps,prd,amssymb,nobibnotes, amsmath,twocolumn,nofootinbib,superscriptaddress,floatfix,preprintnumbers]{revtex4-1}

\usepackage{graphicx,color}
\usepackage{dcolumn}
\usepackage{bm}
\usepackage{hyperref}
\usepackage{amsmath}
\usepackage{mathtools}
\usepackage{multirow}
\usepackage{stackrel}
\usepackage[lofdepth,lotdepth,caption=false]{subfig}

\usepackage{xcolor}

\begin{document}

\begin{flushleft}
LAPTH-039/22
\end{flushleft}

\title{Ultra-high energy neutrinos from high-redshift electromagnetic cascades}

\author{AmirFarzan Esmaeili}
\email{a.farzan.1993@aluno.puc-rio.br}
\affiliation{Departamento de Física, Pontifícia Universidade Católica do Rio de Janeiro, Rio de Janeiro 22452-970, Brazil}

\author{Antonio Capanema}
\email{antoniogalvao@aluno.puc-rio.br}
\affiliation{Departamento de Física, Pontifícia Universidade Católica do Rio de Janeiro, Rio de Janeiro 22452-970, Brazil}

\author{Arman Esmaili}
\email{arman@puc-rio.br}
\affiliation{Departamento de Física, Pontifícia Universidade Católica do Rio de Janeiro, Rio de Janeiro 22452-970, Brazil}
\affiliation{LAPTh, CNRS, USMB, F-74940 Annecy, France}

\author{Pasquale Dario~Serpico}
\email{serpico@lapth.cnrs.fr}
\affiliation{LAPTh, CNRS, USMB, F-74940 Annecy, France}

\begin{abstract}
We study the impact of the muon pair production and double pair production processes induced by ultra-high energy photons on the cosmic microwave background. Although the muon pair production cross section is smaller than the electron pair production one, the associated energy loss length is comparable or shorter than the latter (followed by inverse Compton in the deep Klein-Nishina regime) at high-redshift, where the effect of the astrophysical radio background is expected to be negligible. By performing a simulation taking into account the details of $e/\gamma$ interactions at high energies, we show that a significant fraction of the electromagnetic energy injected at $E\gtrsim 10^{19}\,$eV at redshift $z\gtrsim 5$ is channeled into neutrinos. The double pair production plays a crucial role in enhancing the multiplicity of muon production in these electromagnetic cascades. The ultra-high energy neutrino spectrum, yet to be detected, can in principle harbour information on ultra-high energy sources in the young universe, either conventional or exotic ones, with weaker constraints from the diffuse gamma ray flux compared to their low redshift counterparts. 
\end{abstract}

\maketitle

\section{Introduction\label{sec:intro}}

The high-redshift and high-energy universe remains largely mysterious. The hadronic component of ultra-high energy cosmic rays (UHECRs) suffers significant energy losses, besides lacking directionality due to deflections in extragalactic magnetic fields. Photons and electrons undergo fast interactions with the cosmic microwave background (CMB), quickly degrading their energy, eventually contributing only to the diffuse extragalactic background below the TeV. Even gravitational wave signals at high-redshift will have to wait next generation detectors in order to be properly explored~\cite{Kalogera:2019sui}, and linking them to high-energy counterparts in other messengers is far from evident. Neutrinos are thus the most promising messengers, especially after the detection of astrophysical neutrinos by IceCube~\cite{Aartsen:2013bka,Schneider:2019ayi,Stettner:2019tok,Aartsen:2020aqd}, currently up to $\sim {\cal O}(10)$ PeV~\cite{Aartsen:2021glw}, paving the way to explore the remote and violent universe through these elusive particles. The near future improvements in the sensitivity of neutrino telescopes will further promote this exploration, notably  searching for  ultra-high-energy $\nu$'s ($E\gtrsim 10^{17}$\,eV): Their presence is guaranteed at least from interactions of UHECRs with the CMB (so-called {\it cosmogenic neutrinos}~\cite{Berezinsky:1969erk}), but could also originate from yet unknown astrophysical or exotic processes~\cite{Abbasi:2021vjr,Fang:2017cqe,Batista:2019nhf}.

Of particular interest for diagnostics is the interplay between these messengers, especially between the photons and neutrinos. Gamma rays have long been used to constrain ultra-high energy (UHE) neutrinos, see for instance~\cite{Waxman:1998yy,Mannheim:1998wp,Kalashev:2002kx}. In the following, we revisit the link between these particles at high-$E$ and high-$z$, since differing in some peculiar aspects from the standard expectations. In particular, we will argue that electromagnetic cascades drain some significant energy into the neutrino channel, altering the usual expectation for the ratio of energy into neutrinos vs gamma-rays, besides obviously modifying the expected spectra of UHE neutrinos. The fact that the propagation of UHE photons/electrons may be quantitatively different at high-$z$ was studied in~\cite{Kusenko:2000fk,Postma:2001na}, where the key process responsible for the drainage into the neutrino channel was thought to be $e\gamma\to e\mu\mu$. Soon after, this process was reassessed and found of negligible importance in~\cite{Athar:2001sm}. The same article also suggested that the $\gamma\gamma\to\mu\mu$ process may however play a similar role~\footnote{An early mention of this process can also be found in~\cite{Berezinsky:1970nm}.}. This process has been studied within some approximations in~\cite{Wang:2017phf}, where it was concluded that at low redshift $0\leq z<5$, and due to the interplay with the diffuse universal radiation background (URB)~\footnote{Furthermore, the $e^\pm$ in the cascade could also quickly lose energy via synchrotron emission in intergalactic magnetic fields, if these are close to the current upper limits at the nG level.}, only a relatively small fraction of the initial energy of electromagnetic cascades ($\lesssim 10\%$) channels into neutrinos. Similar considerations were also briefly exposed in~\cite{Li:2007ps}. Here we complement these studies by expanding the redshift range and particularly improving on the microphysics of the cascade development at high energies, leading us to somewhat different conclusions.        

The basic idea is that muon pair production (MPP) is non-negligible in the interaction of UHE photons with CMB, where the subsequent decay of muons generates neutrinos. The MPP introduces an important deviation from the course of well-studied electromagnetic cascade of high energy photons/electrons, where it is the chain of electron pair production (EPP) and inverse-Compton scattering (ICS) which leads to the degradation of initial photon/electron energy and the production of a lower-energy photon spectrum. The large inelasticity in both EPP and ICS renders the MPP feasible since effectively the energy loss length in electromagnetic cascade is larger than the interaction length of MPP. The picture is further altered by the role of the double pair production (DPP), which constitutes an important energy-loss channel and is responsible for lowering the drainage of energy via MPP, while at the same time increasing the muon multiplicity in the cascade.

In section~\ref{sec:interactions} we introduce the microphysics ingredients used to describe the aforementioned processes, and justify qualitatively the importance of MPP. A Monte Carlo simulation of processes and the corresponding distributions of energy drainage and neutrino spectrum are given in section~\ref{sec:results}. Section~\ref{sec:conc} is devoted to the discussions and conclusions, summarizing the take-home message and commenting on possible applications. For the ease of reference, well-known analytical formulae of neutrino spectra emitted in muon decay are reported in Appendix~\ref{nuspectra}.

\section{Photon and electron interactions\label{sec:interactions}}

The propagation of UHE photons and electrons in intergalactic space is hindered by the interaction with background photon fields ($\gamma_{\rm b}$) that permeate the Universe. Starting with an UHE photon, the main relevant interactions are EPP ($\gamma\gamma_{\rm b}\to e^+e^-$) and ICS ($e\gamma_{\rm b}\to e\gamma$) where their successive iteration develops the conventional ``electromagnetic cascade". At variance with low-$E$ and low-$z$ cascades,  MPP ($\gamma\gamma_{\rm b}\rightarrow\mu^-\mu^+$) and DPP ($\gamma\gamma_{\rm b}\rightarrow e^+e^-e^+e^-$) are also of interest for our analysis and will significantly contribute to the cascade development. At low-$z$, UHE photons are dominantly interacting with the URB and electrons are possibly affected by syncrhotron emission on extragalactic magnetic fields, if close to the allowed upper limits of nG strength~\cite{Lee:1996fp}. However, the URB is expected to drop at $z\gtrsim 2$, and by $z\sim 5$ it should be vanishingly small compared to the CMB~\cite{Lee:1996fp, Kusenko:2000fk, Postma:2001na}. While the exact redshift at which the URB can be neglected is poorly known, it is robust to assume that such a redshift exists, since the URB is of astrophysical origin and, while the CMB density of photons grows with $z$ as $(1+z)^3$, the density of astrophysical sources drops at $z>2$ and should eventually be vanishing at $z\gtrsim 15$. Based on models in the literature, we estimate that $z=5$ is a rather conservative assumption, hence we will show results considering interactions solely with the CMB only above this redshift. 

The total and differential cross sections of EPP (also dubbed Breit-Wheeler process~\cite{Breit:1934zz})  are respectively  \cite{Berestetskii:1965xsa}
\begin{multline}\label{eq:PP_tot_xsec}
    \sigma_{\rm EPP} = \sigma_{\rm T}\,\frac{3}{16}(1-\beta^2) \bigg[ (3-\beta^4)\ln \frac{1+\beta}{1-\beta} \\
    - 2\beta(2-\beta^2) \bigg]~,
\end{multline}
and
\begin{multline}\label{eq:PP_diff_xsec}
    \frac{{\rm d}\sigma_{\rm EPP}}{{\rm d}E_e} = \sigma_{\rm T}\,\frac{3}{4}\,\frac{m_e^2}{s}\,\frac{1}{E_\gamma} \left[ \frac{E_e}{E_\gamma - E_e} + \frac{E_\gamma - E_e}{E_e} \right. \\ + E_\gamma\left(1-\beta^2\right)\left( \frac{1}{E_e} + \frac{1}{E_\gamma - E_e} \right) \\ \left. - \frac{E_\gamma^2\left(1-\beta^2\right)^2}{4}\left( \frac{1}{E_e} + \frac{1}{E_\gamma - E_e} \right)^2 \right]~,
\end{multline}
where $m_e$ is the electron mass, $\sigma_T=8\pi\alpha^2/(3m_e^2)$ is the Thomson cross section (in the whole paper, natural units are used), $\alpha$ the fine structure constant, $\beta = \sqrt{1-4m_e^2/s}$ is the velocity of outgoing electron in the CM frame, $s =2E_\gamma\epsilon(1-\mu)$ is the squared CM energy, $\epsilon$ and $E_\gamma$ are respectively the energies of the target (here CMB photon) and high energy photons, $\mu$ is the cosine of the angle between the momenta of the incoming photons, and $E_e$ is the energy of the produced electron (or positron) whose  allowed range is $(1-\beta)/2\leq E_e/E_\gamma\leq(1+\beta)/2$.

The ICS total and differential cross sections are given by \cite{Aharonyan:1981csr}
\begin{multline}\label{eq:ICS_tot_xsec}
    \sigma_{\rm ICS} = \sigma_{\rm T}\,\frac{3}{8}\,\frac{m_e^2}{\tilde{s}\tilde{\beta}} \left[\frac{2}{\tilde{\beta}(1+\tilde{\beta})}(2 + 2\tilde{\beta} - \tilde{\beta}^2 - 2\tilde{\beta}^3) \right. \\ \left. - \frac{1}{\tilde{\beta}^2}(2 - 3\tilde{\beta}^2 - \tilde{\beta}^3) \ln \frac{1+\tilde{\beta}}{1-\tilde{\beta}} \right]~,
\end{multline}
and
\begin{multline}\label{eq:ICS_diff_Xsec_CM}
    \frac{{\rm d}\sigma_{\rm ICS}}{{\rm d}E_e'} = \sigma_{\rm T}\,\frac{3}{8}\,\frac{m_e^2}{\tilde{s}}\,\frac{1}{E_e}\,\frac{1+\tilde{\beta}}{\tilde{\beta}} \left[ \frac{E'_e}{E_e} + \frac{E_e}{E'_e} \right. \\ \left. +\frac{2(1-\tilde{\beta})}{\tilde{\beta}} \left( 1 - \frac{E_e}{E'_e} \right) + \frac{1-\tilde{\beta}^2}{\tilde{\beta}^2} \left( 1 - \frac{E_e}{E'_e} \right)^2 \right]~,
\end{multline}
where $\tilde{\beta} =(\tilde{s}-m_e^2)/(\tilde{s}+m_e^2)$ is the velocity of the outgoing electron in the CM frame, $\tilde{s}=m_e^2+2\epsilon(E_e-\mu\sqrt{E_e^2-m_e^2})$ is the squared CM energy, $E_e$ is the energy of the initial high-energy electron, and $E'_e$ is the energy of the outgoing electron, whose allowed range is $(1-\tilde{\beta})/(1+\tilde{\beta})\leq E'_e/E_e\leq 1$.

The threshold energy for EPP derives from the condition $s= 4m_e^2$. For $s\geq 4m_\mu^2$ ($m_\mu$ being the muon mass), MPP ($\gamma\gamma_{\rm b}\rightarrow\mu^-\mu^+$) also becomes accessible. The MPP's total cross section, $\sigma_{\rm MPP}$, and differential cross section, ${\rm d}\sigma_{\rm MPP}/{\rm d}E_\mu$, can be obtained by replacing $m_e\to m_\mu$ in the formulae for EPP.  For the MPP on the bulk of the CMB at redshift $z$, the threshold energy is $E_{\rm MPP}^{\rm thr}=m_\mu^2/\langle\epsilon \rangle\simeq 1.8\times10^{19}/(1+z)$~eV, for the benchmark value $\langle\epsilon \rangle \simeq 2.7 (1+z)T_0$, where $T_0=2.35\times 10^{-4}$~eV is the current CMB temperature. 

At high energies the DPP becomes relevant. While a fully analytic expression for $\sigma_{\rm DPP}$ is quite involved, it rapidly converges to the constant value $\sigma_{\rm DPP}^\infty\simeq 6.45\times 10^{-30}~{\rm cm}^2$ above its threshold at $s=16m_e^2$, and its energy dependence can be closely approximated by $\sigma_{\rm DPP} \approx (1-4m_e^2/E_\gamma \epsilon)^6 \sigma_{\rm DPP}^\infty$ \cite{Brown:1973dpp}. We will not employ the full expression for the DPP differential cross section, but simply approximate the process in assuming that one of the pairs carries the quasi-totality of the projectile photon energy, sharing it equally. This captures the main quantitative effect of DPP on the cascade development~\cite{Demidov:2009dpp}. 

Although MPP is suppressed with respect to EPP (for example, at $s=10^{18}~\rm eV^2$, $\sigma_{\rm MPP}/\sigma_{\rm EPP}\approx0.26$), muon production is definitely relevant since EPP and ICS have large inelasticities at high energies. The inelasticity of a given process, i.e. the average fraction of energy transferred from the initial leading particle to the produced leading particle, is given by
\begin{equation}\label{eq:elasticity}
    \eta(s) = \frac{1}{\sigma(s)}\int {\rm d}E'\,\frac{E'}{E_0}\, \frac{{\rm d}\sigma}{{\rm d}E'}(E',s)~,
\end{equation}
where $E'$ is the energy of the produced leading particle and $E_0$ is the energy of the initial leading particle.
Due to the large inelasticities in EPP and ICS, the initial UHE photon (or electron) undergoes a sequence of EPP+ICS, at each step of which the leading particle emerges with an energy close to the initial one. If MPP happens, this sequence is greatly altered since the final-state $e^\pm$ from muon decay carry a comparatively small fraction of the parent photon energy. For the DPP, we assume that each one of the leptons in a pair $e^+e^-$ carries half of the initial photon energy, which is very close to the actual (and much more involved) calculation~\cite{Demidov:2009dpp}.

To quantify the relative prominence of these processes, let us define the {\it interaction length} (or mean free path) 
\begin{equation}
   \lambda_p(E) = \frac{1}{\int {\rm d}\epsilon \int{\rm d}\mu\, P(\mu)\, n_{\rm CMB}(\epsilon)\,\sigma_p(s)}~,\label{Intlength}
\end{equation}
where $p=$~EPP, MPP or DPP, $n_{\rm CMB}$ is the number density of CMB photons per unit energy, $s=s(E,\epsilon,\mu)$, $P(\mu)=(1-\mu)/2$ is the distribution of collision angles (or flux factor), and the integral over $\mu$ extends up to $1-2m^2/E_\gamma\epsilon$, with $m$ being either $m_e$ or $m_\mu$. Similarly, one can define the {\it energy loss length}~\footnote{A more correct definition in terms of stopping power ${\rm d} E/{\rm d} x$ would be $\Lambda=\int {\rm d} x\, E/(-{\rm d} E/{\rm d} x)$, but for moderate energy dependence of the integrand it leads to comparable results. We content ourselves with the simplest definition of eq.~(\ref{Elosslength}), given the mere illustrative purpose of this quantity in this section.}
\begin{equation}
    \Lambda_p(E) = \frac{1}{\int {\rm d}\epsilon \int{\rm d}\mu\, P(\mu)\, n_{\rm CMB}(\epsilon)\,\sigma_p(s)\,[1-\eta_p(s)]}~.\label{Elosslength}
\end{equation} 
For a catastrophic event like MPP, $\Lambda\simeq \lambda$, but whenever only a small fraction of energy is lost at each interaction, as it is for the leading particle in EPP/ICS cycles, $\Lambda\gg \lambda$.

\begin{figure}[!ht]
    \centering
    \includegraphics[width=0.48\textwidth]{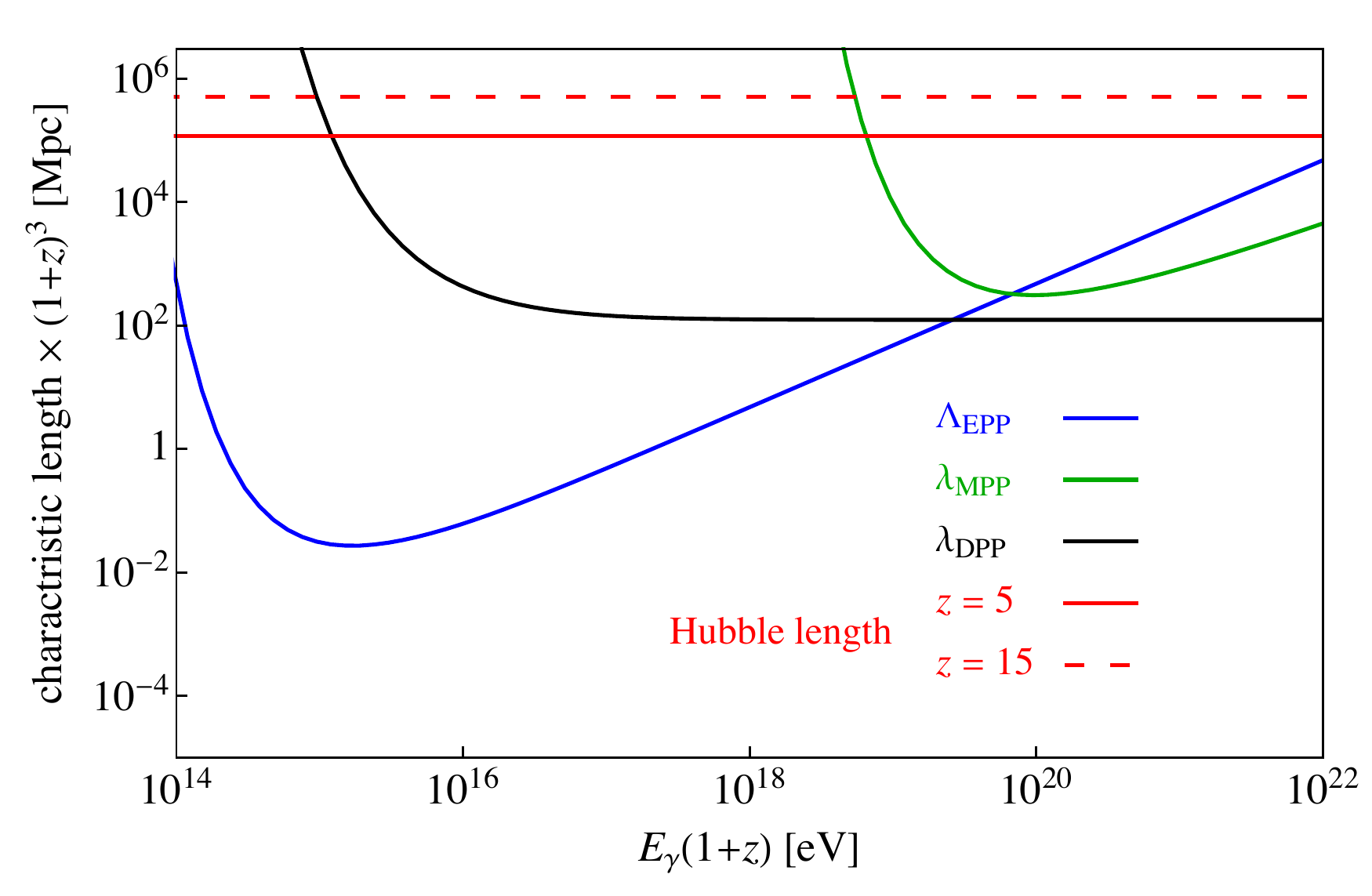}
    \caption{\label{fig:length_comparison}Comparison between $\Lambda_{\rm EPP}$ (blue), $\lambda_{\rm MPP}\simeq \Lambda_{\rm MPP}$ (green), $\lambda_{\rm DPP}$ (black) and Hubble $1/H(z)$ (red) lengths rescaled by $(1+z)^3$ vs. energy of the UHE photon, rescaled by $(1+z)$. The Hubble lengths are displayed for redshifts $z=5$ (solid red line) and $z=15$ (dashed red line).}
\end{figure}

Figure~\ref{fig:length_comparison} compares $\Lambda_{\rm EPP}$ (blue color), $\lambda_{\rm MPP}\simeq \Lambda_{\rm MPP}$ (green color) and $\lambda_{\rm DPP}$ (black color) as functions of the UHE photon's energy at the observation point. The vertical and horizontal axes are scaled by $(1+z)^3$ and $(1+z)$, respectively, thus the curves are valid for any redshift \footnote{Since the differential density $n_{\rm CMB}(\epsilon)$ scales as $(1+z)^2$, lengths get contracted by $(1+z)$ and energies increase by $(1+z)$ with $z$, this implies that $\lambda(E,z)=(1+z)^{-3}\,\lambda_0[E(1+z)]$. Ditto for $\Lambda$.}. Remarkably, $\lambda_{\rm MPP} < \Lambda_{\rm EPP}$ at high energies indicates that we expect MPP to happen before the photons/electrons have lost a significant fraction of their initial energy via the EPP/ICS cycle. On the other hand, since $\lambda_{\rm DPP} \lesssim \Lambda_{\rm EPP}$ at high energies, DPP affects the cascade development; similarly, $\lambda_{\rm DPP} < \lambda_{\rm MPP}$ signals the relevance of DPP inclusion in the study of MPP. Qualitatively, starting from a very high energy photon, we expect the role of DPP is to split the initial photon energy into a pair $e^\pm$ almost equally. This is followed by ICS events, where the upscattered  photons initiate a new multiplicative process via DPP and so on, until the photon energies end-up close to the minimum of the MPP interaction length,  around $E_\gamma(1+z)\simeq 10^{20}{\rm eV}$. At that point,  the particles are only a factor $2-3$ less likely to undergo muon generation via MPP than to degrade below MPP threshold via a final DPP event, or to start a ``conventional'' cascade via EPP; this explains why MPP matters. Since in a MPP event about 65\% of the energy is carried by the neutrinos (see Appendix~\ref{nuspectra}), a rough expectation is that, away from threshold effects, on average slightly below $65\%/2\sim 30\%$ of  $E_\gamma$ is drained into the neutrino flux. We also expect that the higher the energy, the larger is the multiplicity of muons through which the drainage is happening, with this number scaling proportionally to $E_\gamma(1+z)/10^{20}{\rm eV}$. Finally, we can anticipate that a significant spread around the average should be present due to the stochastic nature of these events.  

Also note that, as shown in Fig.~\ref{fig:length_comparison}, these interaction lengths are well below the Hubble length~\footnote{$H(z)\simeq H_0\sqrt{\Omega_\Lambda+\Omega_M(1+z)^{3}}$ being the Hubble expansion rate, $H_0\simeq 70~$km/s/Mpc its current value, and $\Omega_M\simeq 0.3$ the matter density of the universe in terms of the critical one. Instead, $\Omega_\Lambda\simeq 1-\Omega_M$ is the dimensionless energy density of the cosmological constant, playing a negligible role at redshifts of interest here. } $H(z)^{-1}$. Hence, particle dynamics rather than cosmology rules the evolution in $E$-space, with the cascade development that can be considered almost instantaneous in $z$.

These qualitative arguments motivate a more quantitative study of the effect of this process, which we embark on in the next section. Before moving to that, let us mention our rationale for ignoring some additional processes (a synoptic description of which can be found in~\cite{Lee:1996fp}). Charged pion production ($\gamma\gamma_{\rm b}\rightarrow \pi^+\pi^-$) becomes possible at $s\geq 4m_{\pi^\pm}^2$, but its cross section is only comparable to EPP and MPP in a small window of energies (corresponding to the $f_2(1270)$ resonance)~\cite{Boyer:1990opp,Morgan:1987cej} and is otherwise sub-leading. Including this process would only mildly strengthen the conclusions of this article. The production of neutral pions, kaons and heavier hadrons in $\gamma\gamma_{\rm b}$ scattering is even more suppressed~\cite{Whalley:2001tpr,Morgan:1994vne}, justifying that we neglect them. Triplet pair production ($e\gamma_{\rm b}\rightarrow e e^+ e^-$), has a cross section comparable to that of ICS already at $\sim 10^{17}$~eV, but leads to a negligible energy loss below $\sim10^{22}$~eV since the energy fraction carried by the produced pair is very small ($\sim 10^{-3}$)~\cite{Sigl:2001wih} and its inclusion is expected to change our conclusions at the few percent level at most. Finally, we also neglect the synchrotron energy losses of UHE electrons onto extragalactic magnetic fields. While these may be of importance at low redshift, see~\cite{Wang:2017phf}, unless the fields are of primordial origin, their role with respect to losses on CMB should vanish going to high-$z$, with an argument qualitatively similar to what we discussed for the URB. Note that, despite limited information on extragalactic magnetogenesis, current evidence suggests indeed that extragalactic fields grow at low-$z$ via an astrophysical dynamo mechanism, rather than being primordial~\cite{Pomakov:2022poi} (or implying much smaller primordial seeds), consistent with the hypothesis done here.

\section{Simulation and results\label{sec:results}}

To quantitatively assess the role of MPP at high energies, we proceed with a Monte Carlo simulation. This is unavoidable if one is to take into account the discrete and stochastic nature of the processes. As previously discussed, it turns out that the mean free path between interactions is so short compared to the cosmological scales that the change in the redshift of two successive processes can be safely ignored. Thus, starting with a photon~\footnote{Starting with an electron would not lead to appreciable differences.} with specified energy $E_\gamma$ and redshift $z$, only the evolution in $E-$space is relevant, described as a sequence of interactions where the leading electromagnetically interacting particle's energy degrades, until the MPP process is no longer kinematically open. At each photon interaction, we compare a random number in [0,1] with the probability to yield a MPP
\begin{equation}\label{eq:pmpp}
    p_{\rm MPP}= \frac{\lambda_{\rm MPP}^{-1}}{\lambda_{\rm EPP}^{-1} + \lambda_{\rm MPP}^{-1} + \lambda_{\rm DPP}^{-1}}\,.
\end{equation}
As an example, at $z=15$ we estimate $p_{\rm MPP}\approx(0.07, 0.07, 0.02)$ respectively for $E_\gamma=(10^{19},10^{20},10^{21})$~eV. The probability of DPP can be defined similarly to Eq.~(\ref{eq:pmpp}) by replacing $\lambda_{\rm MPP}^{-1}$ with $\lambda_{\rm DPP}^{-1}$ in the numerator, leading to $p_{\rm DPP}\approx(0.18, 0.62, 0.93)$ for the same $E_\gamma$'s and $z$. Obviously, for EPP we have $p_{\rm EPP} = 1-p_{\rm DPP}-p_{\rm MPP}$. The cascade development depends on the selected interaction at each step. When MPP is chosen, the $e^\pm$ from the $\mu^\pm$ decay are injected again into the simulation by performing ICS on CMB and creating an UHE photon which starts a new branch of cascade. If DPP is chosen, an $e^+e^-$ pair will be followed (the other pair carrying negligible energy);  each member of the pair, assumed to carry energy $E_\gamma/2$, initiates a new branch after a single ICS event. Finally, the EPP case will be followed by ICS. The photon energy coming from  ICS events, or the  $e^+/e^-$ ones from EPP events are sampled from the corresponding differential cross sections reported in the previous section.   

\begin{figure}[t!]
\centering
\subfloat[]{
\includegraphics[width=0.48\textwidth]{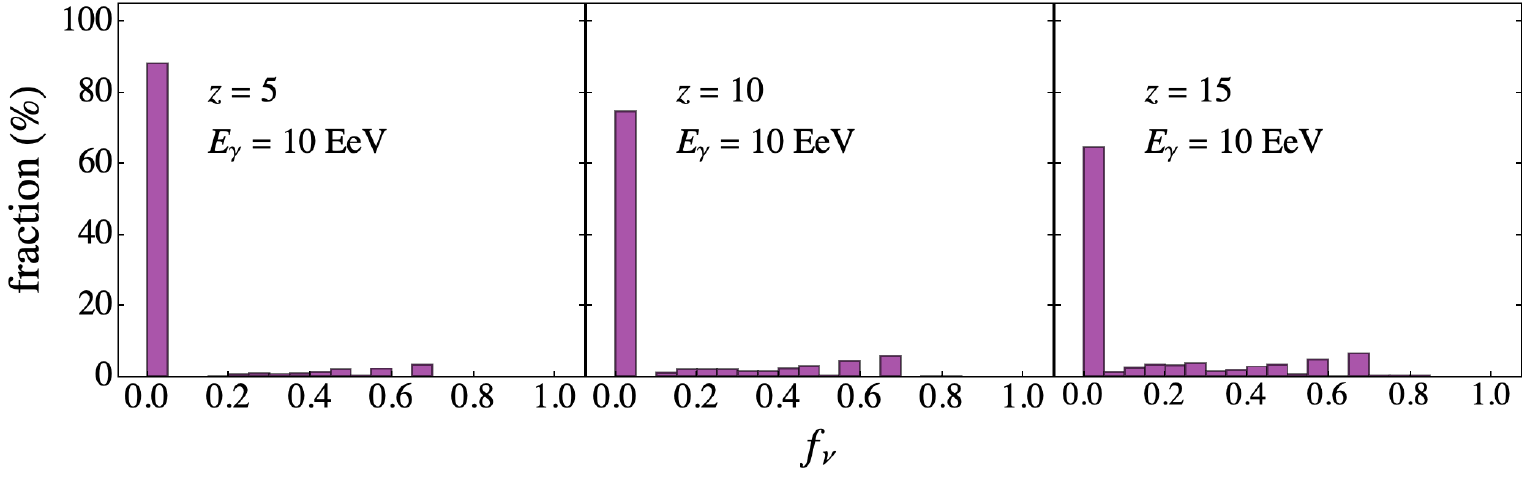}
\label{fig:MCdistE1000}
}
\hspace{0cm}
\subfloat[]{
\includegraphics[width=0.48\textwidth]{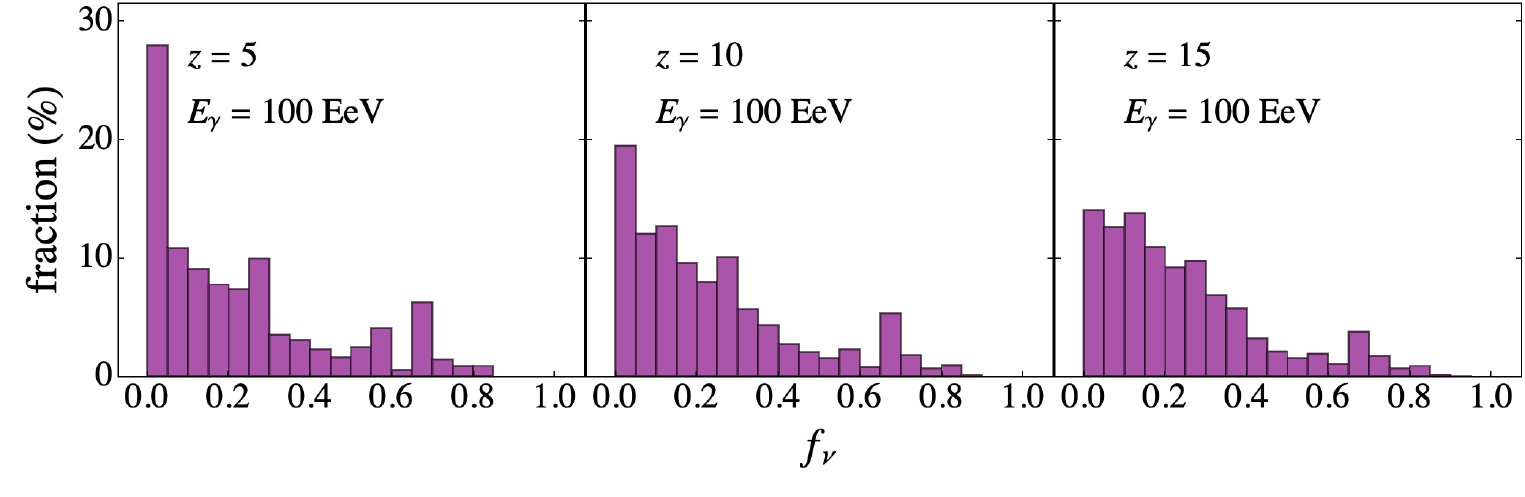}
\label{fig:MCdistE100}
}
\hspace{0cm}
\subfloat[]{
\includegraphics[width=0.48\textwidth]{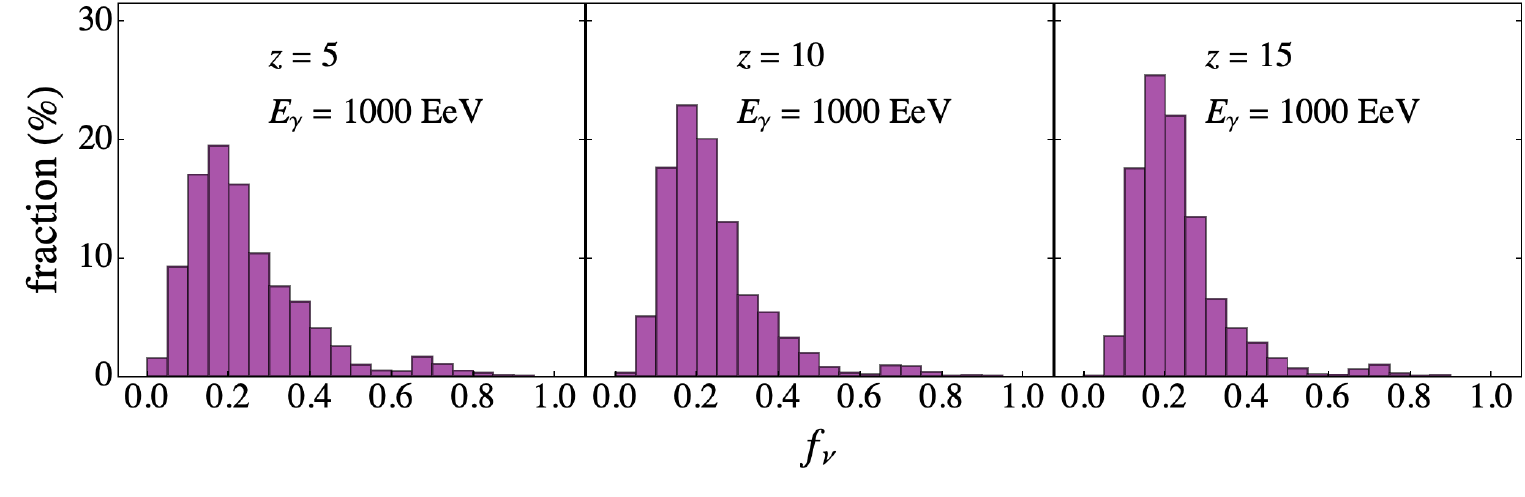}
\label{fig:MCdistE10}
}
\caption{\label{fig:MCdist} The fraction of energy channeling into neutrinos in the simulation of $10^4$ photons with energies $E_\gamma=10^{19}$~eV, $10^{20}$~eV and $10^{21}$~eV, respectively in panels (a), (b) and (c). The left, middle and right panels are respectively for injection at $z=5$, 10 and 15.}
\end{figure}

The quantities of main interest for phenomenology are the fraction of the initial photon energy channeled into neutrinos, {$f_\nu$}, and the neutrino spectra resulting from this process.
Figure~\ref{fig:MCdist} shows the distribution of $f_\nu$ for $10^4$ injected photons with energy $E_\gamma=10^{19}$~eV, $10^{20}$~eV and $10^{21}$~eV in panels (a), (b) and (c), respectively, from top to bottom; the panels in the leftmost column report results at $z=5$, in the middle one at $z=10$  and in the rightmost column at $z=15$. The plots show that at 10$^{19}\,$eV and $z=5$ only about 12\% of the photons experience MPP. This fraction grows to about 25\% at $z=10$ and 35\% at $z=15$. At 10$^{20}\,$eV, well above 70\% of photons experience MPP at $z\geq 5$, with this fraction exceeding 94\% at $z= 15$. Eventually, for $E_\gamma=10^{21}\,$eV, basically every cascade involves one or more MPP events. 
This behaviour makes sense once realising that, at lower energy, threshold effects reducing the importance of MPP are important. At the highest energy, as discussed, the multiplicity of energetic $e^\pm$ due to DPP makes the probability that none of them initiate a MPP vanishingly small. Note how the distributions of $f_\nu$ are  broad (and skewed), reflecting the stochastic nature of the processes.

\begin{figure}
    \centering
        \includegraphics[width=0.48\textwidth]{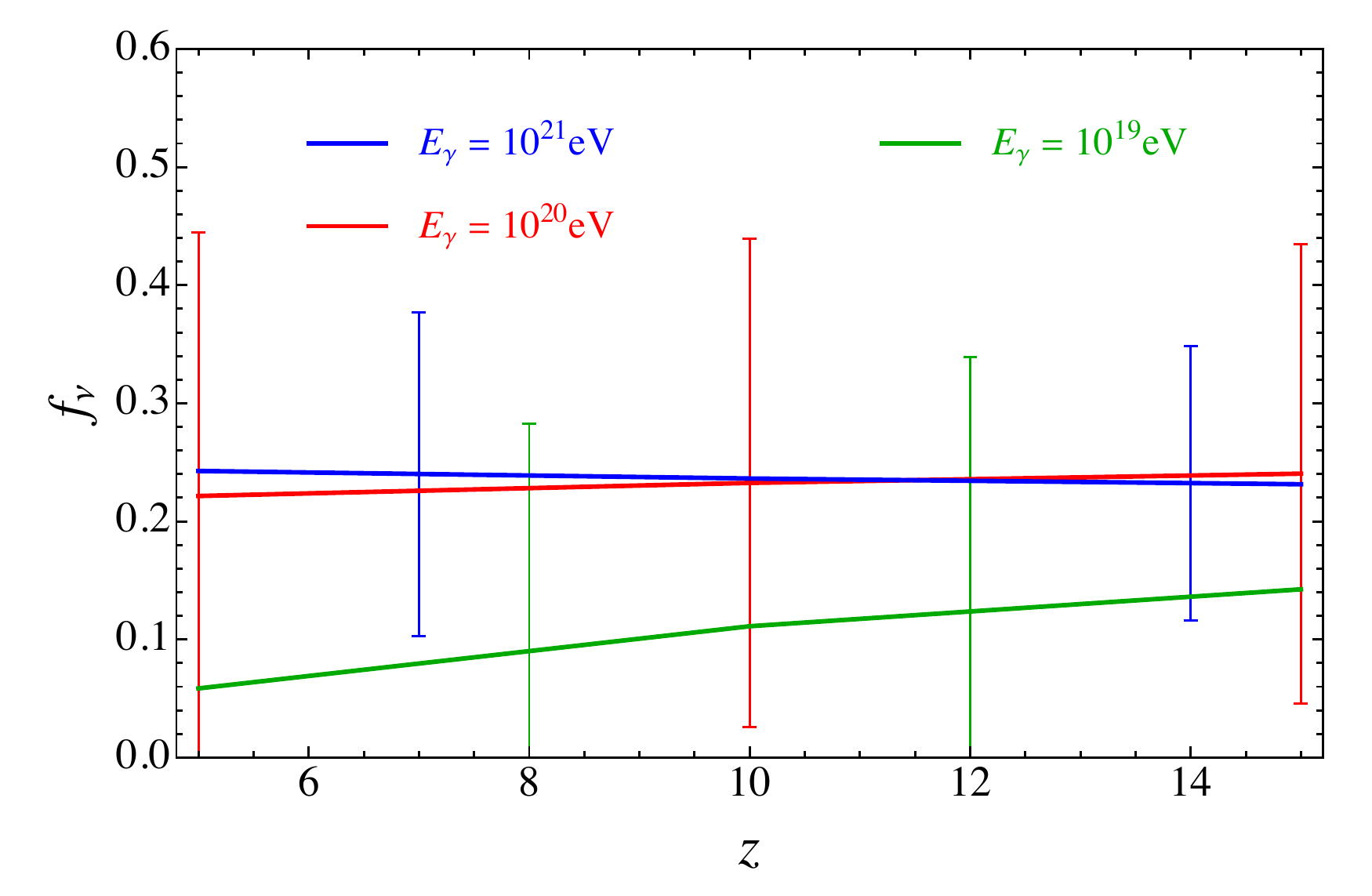}
    \caption{The mean fraction of the initial photon's energy ending up in neutrinos, $f_\nu$, for three different energies of the initial photon. The bars show the standard deviation around the mean value depicted by solid curves.}
    \label{fig:Efrac}
\end{figure}

The mean value of $f_\nu$ is a strongly dependent function of energy near the threshold, while being almost constant with energy at high-$E$, as reported in Figure~\ref{fig:Efrac}, for the initial photon energies $E_\gamma=10^{19}$~eV (green), $10^{20}$~eV (red), and $10^{21}$~eV (blue). The bar around each curve shows the standard deviation, calculated from the distributions in Figure~\ref{fig:MCdist}. It  mildly shrinks with $E_\gamma$, since high multiplicities make the process ``more deterministic''.

\begin{figure}[t!]
\centering
\subfloat[]{
\includegraphics[width=0.48\textwidth]{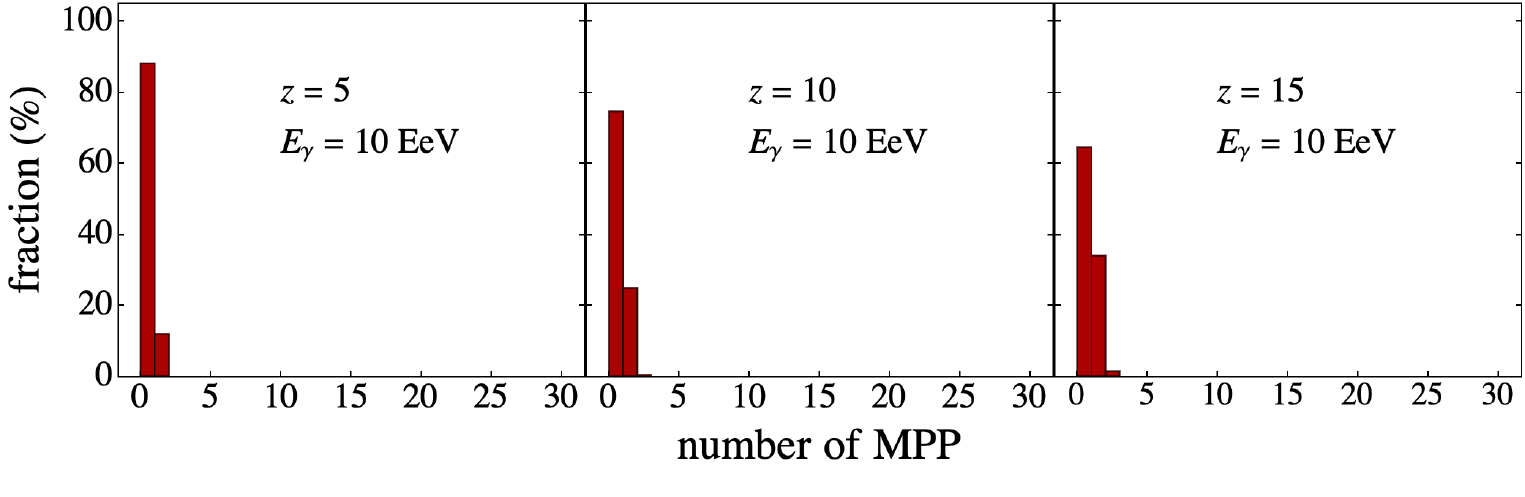}
\label{fig:MPPnumberDistE1000}
}
\hspace{0cm}
\subfloat[]{
\includegraphics[width=0.48\textwidth]{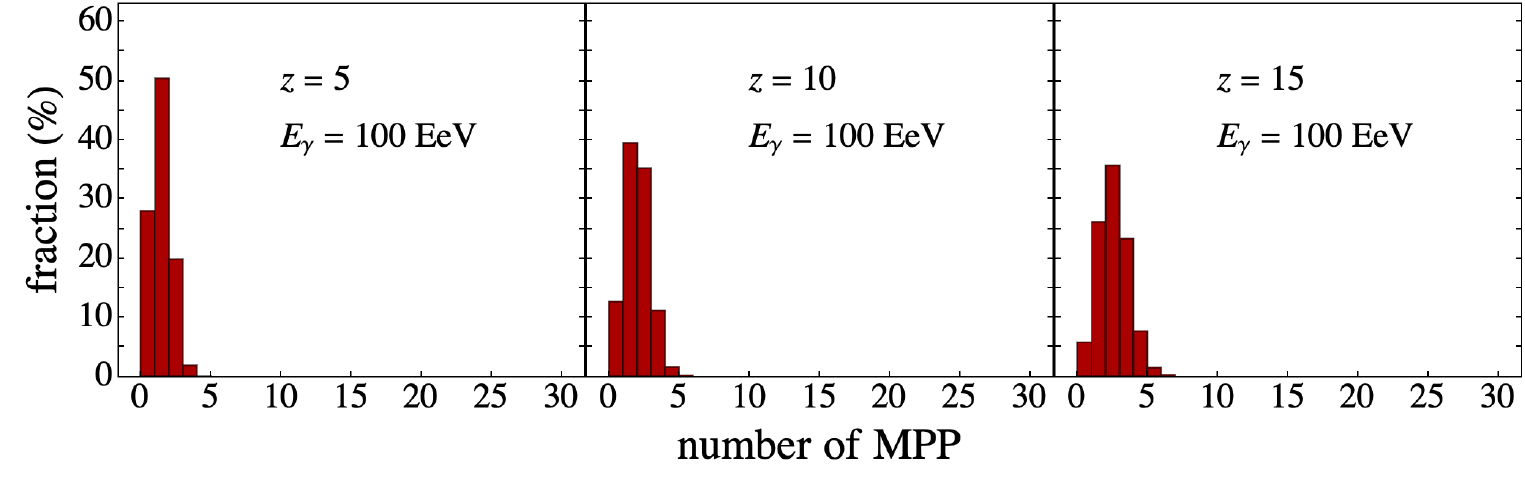}
\label{fig:MPPnumberDistE100}
}
\hspace{0cm}
\subfloat[]{
\includegraphics[width=0.48\textwidth]{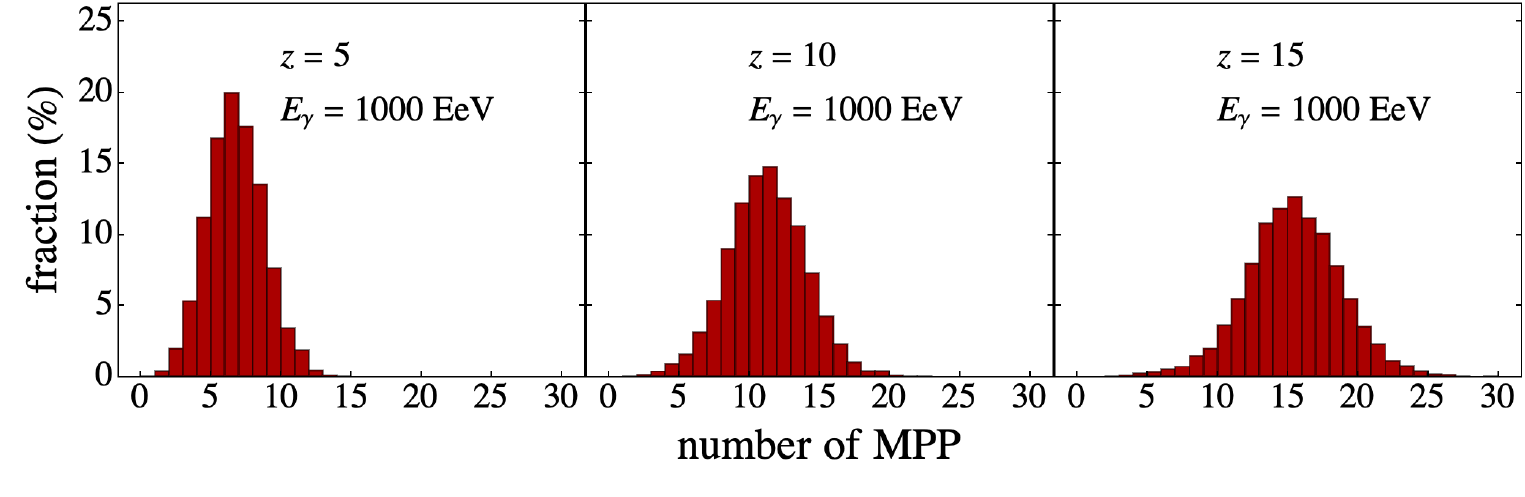}
\label{fig:MPPnumberDistE10}
}
\caption{\label{fig:MPPnumberDist} The distribution of the number of MPP occurrence, for the energies and redshifts as in Figure~\ref{fig:MCdist}.}
\end{figure}

Figure~\ref{fig:MPPnumberDist} illustrates the point that, especially at high-$E_\gamma$ and high-$z$, the multiplicity of muons via MPP events is considerable, for the reasons described in the previous section. For instance, at $E_\gamma=10^{21}$~eV, on average $\sim 6,11$ and 15 MPPs will be realized for injection at $z=5,10$ and 15, respectively. Even if the MPP process typically intervenes only when the particles have degraded to energies significantly lower than the injected ones, its multiplicity makes its impact on the energy budget not negligible. Note that in the early study~\cite{Li:2007ps} this aspect was completely missed ``by construction'', since no follow-up of the leptons produced via DPP was performed. Their estimate of only $\sim$~few percent of the electromagnetic energy drainage into neutrinos is thus not only due to the different conditions relevant at low-$z$, but also to the fact that they did not include this important effect.

\begin{figure}[t!]
    \centering
    \includegraphics[width=0.48\textwidth]{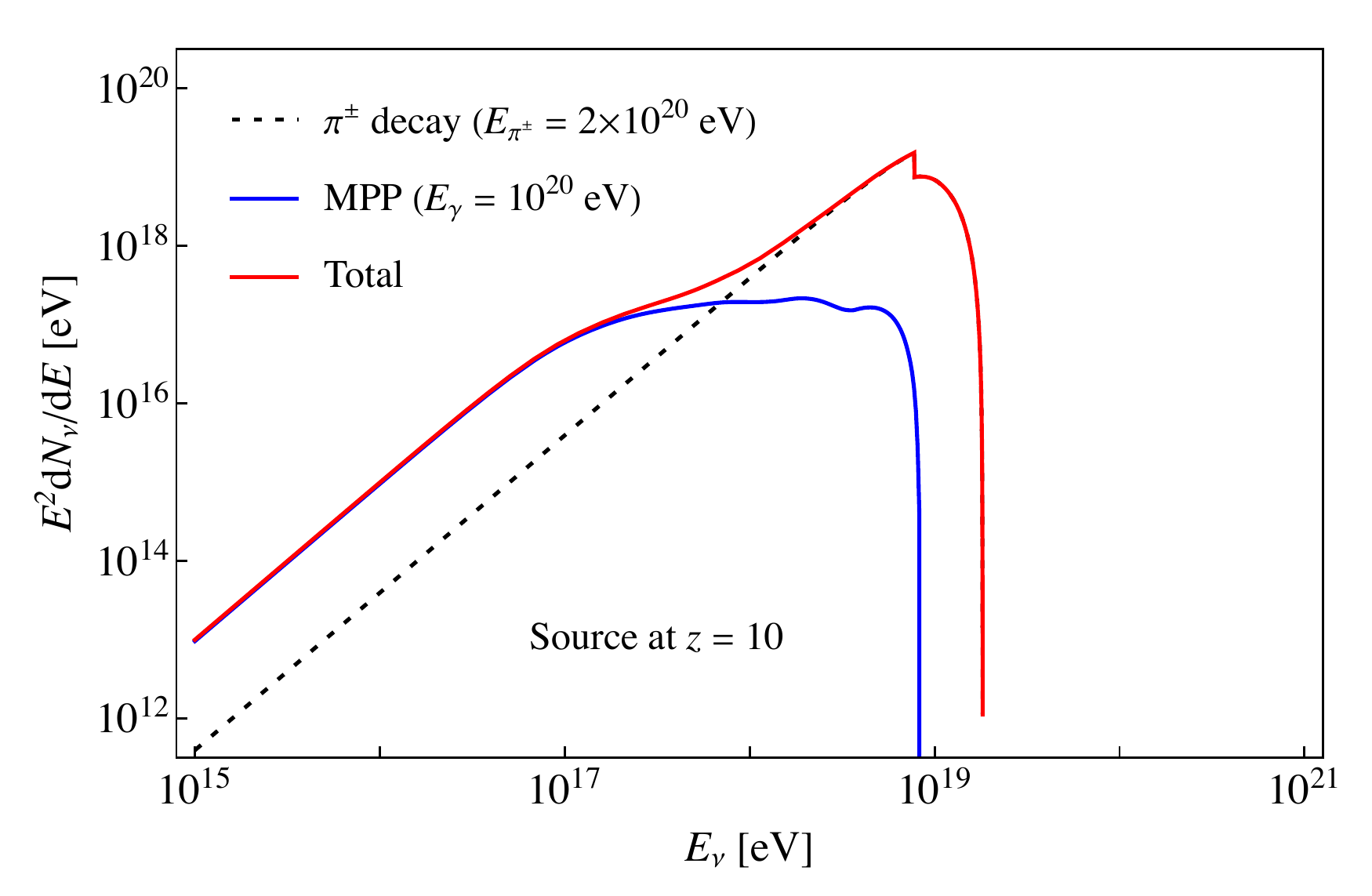}
    \caption{Neutrino spectrum {\it at the Earth} from MPP (solid blue curve, this work) and charged pions decay chain (dotted black curve), from a source at redshift $z=10$ injecting photons at $E_\gamma=10^{20}$~eV. The solid red curve is their sum.}
    \label{fig:nuSpec}
\end{figure}

The average all-flavor neutrino spectrum {\it at the  Earth} from a photon injected at $z=10$ with energy $E_\gamma=10^{20}$~eV is depicted in Figure~\ref{fig:nuSpec} by the blue solid curve. This is based on the well-known analytical descriptions of neutrino spectra from muon decay (see the formulae in~\cite{Lipari:1992lse}, also summarised in Appendix~\ref{nuspectra}) which have been averaged over the $10^4$ injected photons in our simulation. The wiggles at the peak come from multiple MPPs which for the case of Figure~\ref{fig:nuSpec} can be up to five MPPs, with $\sim 46\%$ of cases leading to two or three MPPs. In a conventional scenario, UHE photons are the product of decays of $\pi^0$'s, that are unavoidably accompanied by $\pi^\pm$'s, whose decays produce neutrinos. In Figure~\ref{fig:nuSpec} we also show, by the black dotted curve, the neutrino spectrum from the decay chain $\pi^\pm \to \mu^\pm\nu_\mu(\bar{\nu}_\mu)\to e^\pm \nu_\mu\bar{\nu}_\mu \nu_e (\bar{\nu}_e)$ (see the formulae in Appendix~\ref{nuspectra}) with the energy of $\pi^\pm$ equal to $2\times10^{20}$~eV. The little discontinuity in the dotted curve comes from the contribution of the neutrino emitted directly from the pion decay  $\pi^\pm \to \mu^\pm\nu_\mu(\bar{\nu}_\mu)$. Note how the neutrinos from MPP emerge over those from $\pi^\pm$ in the low-energy part of the distribution, where they dominate the flux by one order of  magnitude.  

\begin{figure}[t!]
    \centering
    \includegraphics[width=0.48\textwidth]{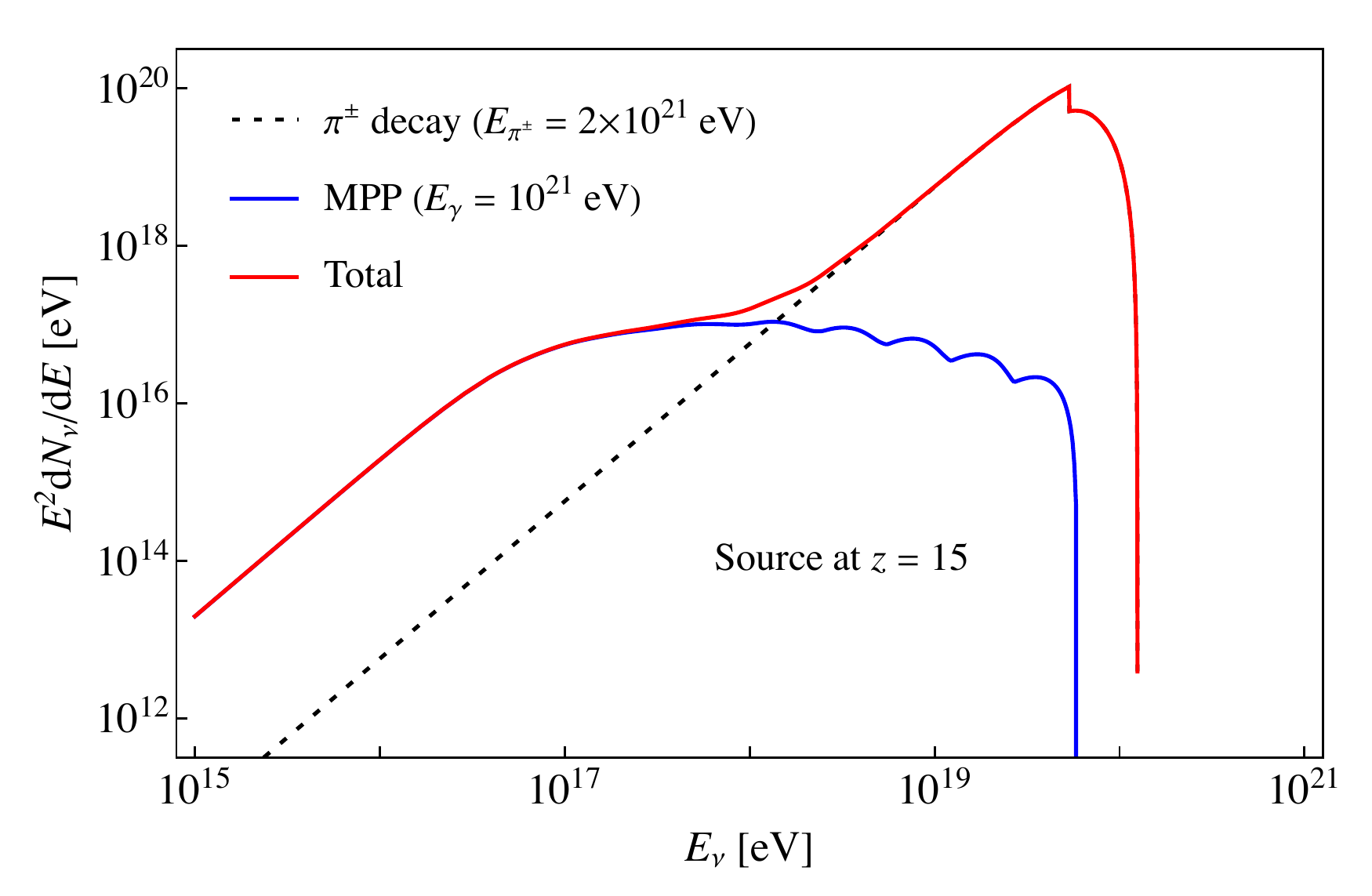}
    \caption{As Fig.~\ref{fig:nuSpec}, for a source at redshift $z=15$ injecting photons at $E_\gamma=10^{21}$~eV}
    \label{fig:nuSpec-2}
\end{figure}

At higher $E_\gamma$ and $z$, where the number of MPPs grows, yet more pronounced features are expected in the neutrino spectrum, as can be seen in Figure~\ref{fig:nuSpec-2} which shows the case of $E_\gamma=10^{21}$~eV injected at $z=15$. The same features of Figure~\ref{fig:nuSpec} are now present in a more exacerbated form. This clearly illustrates the relevance of the MPP process in shaping the UHE neutrino spectra from high-$z$/high-$E$ sources.

Another implication worth commenting upon is that the process discussed here alters the multimessenger $\gamma$-$\nu$ correlation. Conventional production scenarios arising from $pp$ or $p\gamma$ interactions in an UHE astrophysical source predict that the neutrino and gamma-ray emission spectra are related by
\begin{equation}\label{eq:nugamma}
\varepsilon_\gamma\, \frac{{\rm d}N_\gamma}{{\rm d}\varepsilon_\gamma} = \frac{1}{3K_\pi} \left[\varepsilon_\nu\, \frac{{\rm d}N_\nu}{{\rm d}\varepsilon_\nu}\right]_{\varepsilon_\nu = \varepsilon_\gamma/2}~,
\end{equation}
where $K_\pi\approx 1\,(1/2)$ is the charged to neutral pion ratio in the $pp\,(p\gamma)$ process. Integrating both sides of eq. (\ref{eq:nugamma}) over energy implies that, at the source, the ratio of  total energies in $\gamma$'s and $\nu$'s obeys ${\cal E}_\gamma \simeq 2/3 \,(4/3)\,{\cal E}_\nu$. The net effect of MPP is to alter this ratio towards the neutrino sector: For example, from Figure \ref{fig:Efrac} we read that for a source emitting $10^{21}$~eV gammas/neutrinos at $z=5$, approximately $24\%$ of the initial photon energy is transferred to neutrinos during the cascade above MPP threshold; naively, the new balance would be ${\cal E}_\gamma' \to 0.76 {\cal E}_\gamma$, ${\cal E}_\nu' \to  \left(1+0.24\times 2/3 \,(4/3)\right) {\cal E}_\nu$, hence ${\cal E}_\gamma'/{\cal E}_\nu'\simeq 0.44\, (0.77)$, i.e. a ratio changed by about 40\%. The actual energy budget ending up in the low energy diffuse photon flux is more complicated to compute, since one must account for the contribution seeded by $e^\pm$ from charged pion decays, as well as the fraction of the electromagnetic cascade channelled away by $e^\pm$. However, this simple calculation shows that the role of MPP is to make UHE sources at high redshift {\it darker} than their low-$z$ counterparts even in indirect electromagnetic signals, while making them correspondingly {\it brighter} in the UHE neutrino signal.

\section{Discussion and conclusions\label{sec:conc}}

Despite ongoing searches via instruments like IceCube~\cite{IceCube:2015gsk} or the Pierre Auger Observatory~\cite{PierreAuger:2015ihf}, ultra-high energy neutrinos remain elusive. However, the existence of a guaranteed neutrino flux of cosmogenic origin around $E\sim 10^{17}\,$eV and of viable technology to measure it has stimulated a number of designs to achieve this goal. Even in absence of an earlier serendipitous discovery, an instrument like GRAND~\cite{GRAND:2018iaj} would eventually open up this window in neutrino astronomy, and stimulate further questions. 

In this article, we studied some microphysics aspects associated to the UHE $\nu$ flux production which has been largely neglected: The role of muon pair production in draining energy from electromagnetic messengers at UHE and altering the UHE $\nu$ flux, and its interplay with double pair production. We argued that MPP is expected to be relevant at $E\gtrsim 10^{19}\,$eV and at high redshift $z\gtrsim 5$. The resulting flux would fall at the Earth in the $E\gtrsim 10^{17}\,$eV range of interest for an instrument like GRAND. The physics of the process would somewhat loosen constraints from the diffuse gamma ray flux, and spectral features in the neutrino flux, such as the transition between muon and pion channel production around $10^{18}\,$eV, visible in Figures~\ref{fig:nuSpec} and \ref{fig:nuSpec-2}, may be the least elusive of their signatures. 

We have limited ourselves to generic considerations, to be as model-independent as possible, and  we did not attempt to link the electromagnetic particles injected to primary UHECR fluxes, either.  A few qualitative comments can be however made: 
Naive expectations from energetics would suggest that the UHE neutrino signal is dominated by relatively low-$z$ sources, making hard in this case to dig such a signal out of a larger flux. However, so little is known about UHECR sources at high-$z$, since energy losses make their flux subleading to the low-$z$ one, that one cannot exclude that new classes of very energetic UHE emitters could be unveiled, for which our considerations are particularly relevant. One conceivable example is provided by the processes associated to the birth and growth of supermassive black holes~\cite{Woods:2018lty}, which are still  unsolved astrophysical problems~\cite{Inayoshi:2019fun}. 
Additionally, if an astrophysical flux of UHECRs is present, it is likely dominated by a light (proton-helium) composition, compared to the local observations indicating a nuclear enriched CR composition above $\sim 10$~EeV \cite{Bellido:2018qmw}. This is due to the declining metallicity at high-$z$, see e.g.~\cite{2010MNRAS.408.2115M}. Also, reaching the highest energies considered here clearly relies on the acceleration mechanism not being limited by energy-losses on the CMB, which impose tighter constraints at high-$z$.
Another flux that would have likely escaped detection at low-$z$ could be due to exotic supermassive relics produced in the early universe, if decaying with a lifetime shorter than the Hubble time. Similar scenarios were considered in the past as ``top-down'' models of UHECRs~\cite{Kachelriess:2008bk} and are still considered in relation to dark matter candidates~\cite{Guepin:2021ljb}. Related models would generally leave a major imprint in the UHE neutrino flux,  although the complementary sensitivity of cosmological probes remains to be studied. For decays into hadronic final states, the prompt neutrinos and gamma-rays are expected to originate from comparable numbers of $\pi^+,\pi^0, \pi^-$, so that the spectral considerations made in the previous section should roughly apply. A slightly more favourable situation (in the sense of energetically enhancing the relevance of the effects discussed here) could arise in models with preferentially leptonic final states, with comparable energy budgets of prompt neutrinos (now leading to a quasi-monochromatic spectrum, modulo the $z$-dependence of the source and $Z,W-$strahlung corrections) and charged leptons. Once again, the modified electromagnetic cascades would be visible in the lower energy part of the spectrum. 
 Of course, for definite scenarios which our results apply to, it would be interesting to perform specific calculations, perhaps including also sub-leading microphysics processes, and moving to full-fledged multimessenger studies. Such tasks are left for future investigations. 

In conclusion, if the past is of any guidance, it is wise to be ready to possible surprises from the opening of any new astrophysical window. For the UHE sky at high-$z$, one should be aware that differences are present with respect to naive expectations valid at low-$z$, which is without doubts the most important message of this article. 

\begin{acknowledgments}
A.F.~E. thanks the support received by the Conselho Nacional de Desenvolvimento Científico e Tecnológico (CNPq) scholarship No. 140315/2022-5 and by the Coordenação de Aperfeiçoamento de Pessoal de Nível Superior (CAPES)/Programa de Excelência Acadêmica (PROEX) scholarship No. 88887.617120/2021-00. A.~C. thanks the support received by the CNPq scholarship No. 140316/2021-3 and by the CAPES/PROEX scholarship No. 88887.511843/2020-00. During his stay at Laboratoire d'Annecy-Le-Vieux de Physique Theorique (LAPTh), A.~E. has been partially supported by the Centre national de la recherche scientifique (CNRS) invited researcher program. A.~E. would like to acknowledge support from the International Centre for Theoretical Physics (ICTP) through the Associates Programme (2018-2023). A.~E. thanks partial financial support by the Brazilian funding agency CNPq (grant 407149/2021).
\end{acknowledgments}


\appendix
\section{Neutrino spectra}\label{nuspectra}
Muon decay generates a neutrino spectrum whose shape depends on the energy distribution of produced $\mu^+$ and $\mu^-$; for a monoenergetic UHE photon, their distribution is given by ${\rm d}N_{\mu^\pm}/{\rm d}E_{\mu^\pm}\equiv (1/\sigma_{\rm MPP}){\rm d}\sigma_{\rm MPP}/{\rm d}E_{\mu^\pm}$. The total (all flavors) neutrino spectrum ${\rm d}N_\nu/{\rm d}E_\nu$ from the decay of $\mu^\pm$ with spectrum ${\rm d}N_{\mu^\pm}/{\rm d}E_{\mu^\pm}$ can be written as~\cite{Lipari:1992lse}
\begin{multline}\label{eq:Convolution}
    \frac{{\rm d}N_\nu}{{\rm d}E_\nu}(E_\nu)=\int_{E_{\mu,\rm min}}^{E_{\mu,\rm max}}{\rm d}E_\mu\,\frac{{\rm d}N_{\mu^\pm}}{{\rm d}E_{\mu^\pm}}(E_{\mu}) \\ \times \left[F_{\mu^{\pm}\rightarrow\overset{(-)}{\nu}_\mu}(E_\mu;E_\nu)+F_{\mu^{\pm}\rightarrow\overset{(-)}{\nu}_e}(E_\mu;E_\nu)\right]~,
\end{multline}
where 
\begin{equation}\label{eq:LipariNuFlux3}
    F_{a\rightarrow b}(E_a;E_b)=\frac{1}{E_a}F_{a\rightarrow b}\left(\frac{E_b}{E_a}\right)~,
\end{equation}
and for unpolarized muons one has
\begin{equation}\label{eq:LipariNuFlux1}
    F_{\mu^{\pm}\rightarrow\overset{(-)}{\nu}_\mu}(y)=\frac{5}{3}-3y^2+\frac{4}{3}y^3~,
\end{equation}
and
\begin{equation}\label{eq:LipariNuFlux2}
    F_{\mu^{\pm}\rightarrow\overset{(-)}{\nu}_e}(y)=2-6y^2+4y^3~.
\end{equation}

From these relations we can estimate the total energy drainage from the initial photon to neutrinos. By a simple inspection of the above formulae, on the average $\sim 65\%$ of the energy of a photon at the time of MPP goes to neutrinos.

In Figures \ref{fig:nuSpec} and \ref{fig:nuSpec-2}, we also show the neutrino spectra from a monoenergetic charged pion decay. Given a pion spectrum ${\rm d}N_{\pi^\pm}/{\rm d}E_{\pi^\pm}$, there are two contributions to the final neutrino flux: (i) the (anti)muon neutrino emitted directly from the pion decay $\pi^\pm \to \mu^\pm\nu_\mu(\bar{\nu}_\mu)$,
\begin{equation}
    \frac{{\rm d}N_\nu}{{\rm d}E_\nu}(E_\nu)=\int_{E_{\pi,\rm min}}^{E_{\pi,\rm max}} {\rm d}E_\pi\,\frac{{\rm d}N_{\pi^\pm}}{{\rm d}E_{\pi^\pm}}\,F_{\pi^\pm \rightarrow\overset{(-)}{\nu}_\mu}(E_\pi;E_\nu)~,
\end{equation}
where 
\begin{equation}
    F_{\pi^\pm \rightarrow\overset{(-)}{\nu}_\mu}(x) = \frac{1}{1-r_\pi}[1-\theta(x-1+r_\pi)]~,
\end{equation}
obeys the scaling (\ref{eq:LipariNuFlux3}) and $r_\pi = (m_\mu/m_\pi)^2$, and (ii) the neutrinos emitted in the subsequent muon decay. The latter can be obtained by convoluting
\begin{equation}
    \frac{{\rm d}N_{\mu^\pm}}{{\rm d}E_{\mu^\pm}}(E_{\mu})= \int_{E_{\pi,\rm min}}^{E_{\pi,\rm max}} {\rm d}E_\pi\,\frac{{\rm d}N_{\pi^\pm}}{{\rm d}E_{\pi^\pm}}\,F_{\pi^\pm \rightarrow\mu^\pm}(E_\pi;E_\mu)~,
\end{equation}
in Eq.~(\ref{eq:Convolution}), where
\begin{equation}
    F_{\pi^\pm \rightarrow\mu^\pm}(x)=\frac{1}{1-r_\pi}\theta(x-r_\pi)~.
\end{equation}

\bibliography{references}
\end{document}